\documentclass[preprint,aps,12pt,notitlepage,nofootinbib,tightenlines]{revtex4}
\usepackage{fancyvrb}
\usepackage{booktabs}
\usepackage{amsmath}
\usepackage{accsupp}
\usepackage{graphicx}
\usepackage{enumitem}
\usepackage{underscore}
\usepackage{bm}
\usepackage{times}
\usepackage{braket}
\usepackage{color}
\usepackage{epsfig}
\usepackage{slashed}
\usepackage{hyperref}
\usepackage{multirow}
\usepackage{booktabs}
\usepackage{array}
\usepackage{float}
\usepackage[utf8]{inputenc}
\usepackage[T1]{fontenc}
\newcommand{\beq}{\begin{eqnarray}}
\newcommand{\eeq}{\end{eqnarray}}
\newcommand{\be}{\begin{equation}\begin{aligned}}
\newcommand{\ee}{\end{aligned}\end{equation}}

\definecolor{Red}{rgb}{1.,0.,0.}

\definecolor{Blue}{rgb}{0.,0.,1.}

\definecolor{nicered}{rgb}{0.7,0.1,0.1}
\definecolor{nicegreen}{rgb}{0.1,0.5,0.1}
\def\lsim{ {\ \lower-1.2pt\vbox{\hbox{\rlap{$<$}\lower6pt\vbox{\hbox{$\sim$}}}}\ } }
\def\gsim{ {\ \lower-1.2pt\vbox{\hbox{\rlap{$>$}\lower6pt\vbox{\hbox{$\sim$}}}}\ } }

\hypersetup{colorlinks,citecolor=nicegreen,linkcolor=nicered}
\begin{document}
\title{Probing TeV-scale Majorana neutrinos at a muon-proton collider via a lepton-number-violating fat-jet signature}
\author{Yao-Bei Liu$^{1,2}$\footnote{E-mail: liuyaobei@hist.edu.cn} }
\affiliation{1. Henan Institute of Science and Technology, Xinxiang 453003, China\\
2. School of Electro-Mechanical Engineering, Zhongyuan University of Science and Technology, Xuchang 461000, China\\}

\begin{abstract}
We investigate the sensitivity to TeV-scale heavy Majorana neutrinos at a future muon-proton collider with $\sqrt{s}=5.3$~TeV, based on 1~TeV muon and 7~TeV proton beams. We focus on the lepton-number-violating (LNV) process
$\mu^- p \to jN$, followed by $N\to\mu^+W^-\to\mu^+J$, leading to the characteristic final state $j\mu^+J$. Here, $j$ denotes the associated light jet from heavy-neutrino production, while $J$ denotes the fat jet originating from the hadronic decay of the highly boosted $W$ boson. This topology provides a clean experimental signature with strongly suppressed Standard Model backgrounds. A comprehensive signal and background analysis is performed using parton showering, fast detector simulation, and dedicated fat-jet reconstruction. We derive the projected $2\sigma$ exclusion limits on the muon-flavor mixing parameter $|V_{\mu N}|^2$ for integrated luminosities of $100$~fb$^{-1}$ and $1$~ab$^{-1}$. For heavy-neutrino masses in the range $1~\mathrm{TeV}\leq m_N\leq3~\mathrm{TeV}$, the projected upper limits on $|V_{\mu N}|^2$ range from $2.6\times10^{-6}$ to $5.9\times10^{-5}$ at $1$~ab$^{-1}$. These results show that a future $\mu p$ collider can provide competitive and complementary sensitivity to TeV-scale heavy Majorana neutrinos compared with other proposed high-energy facilities.
\end{abstract}

\maketitle

\newpage
\section{Introduction}
Heavy Majorana neutrinos arise naturally in various beyond-Standard Model (BSM) scenarios, especially those with seesaw mechanisms that explain tiny neutrino masses~\cite{Minkowski:1977sc,Yanagida:1979as,Gell-Mann:1979vob,Glashow:1979nm,Mohapatra:1979ia}. In the minimal type-I seesaw, gauge-singlet right-handed neutrinos $N_R$ are introduced. After electroweak symmetry breaking, left-handed neutrinos $\nu_L$ mix with $N_R$, yielding heavy mass eigenstates $N$ that couple weakly to the SM sector. Their masses may span a broad range, from eV to $10^{14}$~GeV, depending on the model details. The Majorana mass term $\overline{(N_R)^c} M_R N_R$ violates lepton number by two units ($\Delta L = 2$). Thus, searching for lepton-number-violating (LNV) processes provides a direct way to test the Majorana nature of neutrinos.

Neutrinoless double beta decay ($0\nu\beta\beta$) remains the most sensitive low-energy probe of the Majorana nature of neutrinos~\cite{Agostini:2022zub}. At high energies, colliders offer a unique avenue to probe the neutrino mass generation mechanism through clean LNV signatures~\cite{Cai:2017mow}. The production and decay of a heavy neutrino $N$ are governed primarily by its mass $m_N$ and the mixing parameter $|V_{\ell N}|^2$, which characterizes its coupling to a SM neutrino of flavor $\ell$. Experimental constraints are conventionally presented in the $|V_{\ell N}|^{2}$--$m_N$ plane. However, canonical seesaw scenarios generically predict much heavier right-handed neutrinos, with masses extending up to $\mathcal{O}(10^{14}\,\text{GeV})$, a scale far beyond the reach of any terrestrial accelerator. In this high-mass window, cosmological measurements of CMB non-Gaussianity, parametrized by $f_{\rm NL}$, provide a powerful complementary probe~\cite{Han:2024qbw,You:2024hit}.

The LNV process $\ell^- p \to \ell^+ jjj$ at lepton-proton colliders provides a clean probe of heavy Majorana neutrinos and has been investigated extensively, from early studies at HERA~\cite{Buchmuller:1991tu,Buchmuller:1992wm} to more recent systematic analyses for both electron-proton and muon-proton collisions~\cite{Blaksley:2011ey}. Heavy neutrino production has also been studied at the LHC, both by experimental collaborations~\cite{CMS:2018iaf,ATLAS:2019kpx,CMS:2019lwf,CMS:2018jxx,ATLAS:2018dcj,CMS:2022hvh,CMS:2024bni,CMS:2024xdq} and in phenomenological works~\cite{Atre:2009rg,Dev:2013wba,Helo:2013esa,Das:2015toa,Antusch:2018bgr,Babu:2022ycv,Bolton:2024thn,Saha:2026wgv,Pascoli:2018heg}. Future prospects have also been explored at proposed $e^+e^-$ colliders~\cite{Zhang:2018rtr,Mekala:2022cmm,Das:2023tna,Wang:2016eln,Brdar:2025iua,Bellagamba:2025xpd,Liu:2025akp}, muon colliders~\cite{Lu:2022wsm,Mekala:2023diu,Kwok:2023dck,Li:2023tbx,Yang:2023ojm,Jiang:2023mte,Li:2023lkl,Wang:2023zhh,He:2024dwh,Cao:2024rzb}, and $ep$ colliders~\cite{Liang:2010gm,Duarte:2014zea,Mondal:2016kof,Antusch:2016ejd,Lindner:2016lxq,Duarte:2018xst,Li:2018wut,Das:2018usr,Antusch:2019eiz,Azuelos:2021ese,Das:2020gnt,Gu:2022muc,Gu:2022nlj,Zapata:2023wsz,Yang:2023ice,Shen:2025ptk}.

In this work, we consider a muon-proton collider with multi-TeV beam energies. This collider option was initially proposed about two decades ago~\cite{Shiltsev:1997pv,Ginzburg:1998yw,Cheung:1997rg,Cheung:1999wy,Carena:2000su} and has been revisited in recent studies~\cite{Kaya:2018kyt,Kaya:2019ecf,Ketenoglu:2022fzo,Dagli:2022idi,Kaya:2022wrc,Akturk:2024evo}. Muon-proton ($\mu p$) colliders provide a unique and complementary platform for exploring BSM physics, offering higher center-of-mass energy and reduced QCD backgrounds compared to hadron and electron-proton colliders. Benefiting from the intrinsic features of muon beams, $\mu p$ collisions exhibit low pile-up and clean experimental signatures, similar to $ep$ colliders but with a much higher energy reach under the same proton beam conditions. Importantly, $\mu p$ colliders enable on-shell production of heavy neutrinos with masses above $100~\text{GeV}$ via $t$-channel $W$ exchange, which is significantly more efficient than off-shell channels at the LHC. These advantages make future $\mu p$ facilities powerful tools for exploring BSM physics in the TeV mass region~\cite{Caliskan:2017meb,Acar:2017eli,Caliskan:2018vep,Caliskan:2018,Alici:2019asv,Ozansoy:2019rmu,Spor:2020rig,Cheung:2021iev,Aydin:2021iky,Gurkanli:2024tfo,Alici:2024eez,Han:2025itd,Benbrik:2026zjv,Sagheer:2026xvh}.

We extend the search for heavy Majorana neutrinos into the TeV mass range $(1~\text{TeV} \leq m_N \leq 3~\text{TeV})$ at a $\mu p$ collider with $\sqrt{s} = 5.3$~TeV. Unlike previous general analyses~\cite{Blaksley:2011ey}, we focus exclusively on the highly boosted regime where the $W$ boson decay products merge into a single fat-jet, providing a dedicated search strategy for TeV-scale neutrinos. This approach enables a more optimized event selection and offers a cleaner probe of the previously unexplored parameter space in muon-proton collisions.

The remainder of this paper is organized as follows. Section~\ref{sec:model} reviews the theoretical framework of heavy Majorana neutrinos. Section~\ref{sec:analysis} presents the collider simulation, event selection, and sensitivity prospects. Section~\ref{sec:summary} summarizes our findings and discusses the complementarity with other future facilities.

\section{The Model\label{sec:model}}
This analysis employs the minimal Type-I seesaw framework~\cite{delAguila:2008cj, Atre:2009rg}, implemented via the $SM\_HeavyN\_LO$ model for Majorana neutrinos~\cite{Alva:2014gxa, Degrande:2016aje}. In this setup, the SM is augmented by three right-handed neutrino singlets $N_{1,2,3}$, which carry no charges under any SM gauge group.
The full Lagrangian takes the form
\begin{equation}
\mathcal{L}
= \mathcal{L}_{\mathrm{SM}} + \mathcal{L}_{N} + \mathcal{L}_{WN\ell} + \mathcal{L}_{ZN\nu} + \mathcal{L}_{HN\nu},
\end{equation}
where $\mathcal{L}_N$ comprises the kinetic and mass contributions of the heavy neutrinos. (Throughout this work, we adopt 4-spinor notation, combining spinors with dotted and undotted indices):
\begin{equation}
\mathcal{L}_{N} = \frac{1}{2} \cdot  \left(  \bar{N}_{k}i\slashed{\partial}N_{k} - m_{N_{k}}\bar{N}_{k}N_{k} \right) \qquad\textnormal{for}\; k = 1,\,2,\,3,
\end{equation}

The couplings of heavy neutrinos to the $W$, $Z$, and Higgs bosons are encoded in $\mathcal{L}_{WN\ell}$, $\mathcal{L}_{ZN\nu}$, and $\mathcal{L}_{HN\nu}$, respectively. Their explicit expressions read
\beq
\mathcal{L}_{WN\ell} &=& - \frac{g}{\sqrt{2}}W^{+}_{\mu} \sum^{3}_{k=1}\sum^{\tau}_{l=e}\bar{N}_{k}V^{*}_{lk}\gamma^{\mu}P_{L}\ell^{-} + \textnormal{ h.c.},\nonumber \\
\mathcal{L}_{ZN\nu} &=& - \frac{g}{2\cos\theta_{W}}Z_{\mu} \sum^{3}_{k=1}\sum^{\tau}_{l=e}\bar{N}_{k}V^{*}_{lk}\gamma^{\mu}P_{L}\nu_{l} + \textnormal{ h.c.},\nonumber \\
\mathcal{L}_{HN\nu} &=& - \frac{gm_{N}}{2m_{W}}h \sum^{3}_{k=1}\sum^{\tau}_{l=e}\bar{N}_{k}V^{*}_{lk}P_{L}\nu_{l} + \textnormal{ h.c.}
\eeq

The partial decay widths for the three dominant channels of the heavy neutrino are
\beq
\Gamma(N \rightarrow \ell W)
 &=& \frac{g^2 }{64 \pi}|V_{\ell N}|^{2}
 \frac{m_N^3}{m_W^2}\Big(1-\frac{m_W^2}{m_N^2}\Big)^{2}\Big(1+2\frac{m_W^2}{m_N^2}\Big) ,
\nonumber \\
\Gamma(N \rightarrow \nu_\ell Z)
 &=& \frac{g^2}{128 \pi} |V_{\ell N}|^{2}
 \frac{m_N^3}{m_W^2}\Big(1-\frac{m_Z^2}{m_N^2}\Big)^{2}\Big(1+2\frac{m_Z^2}{m_N^2}\Big) ,
\nonumber \\
\Gamma(N \rightarrow \nu_\ell h)
 &=& \frac{g^2}{128 \pi} |V_{\ell N}|^2 \frac{m_N^3}{m_{W}^{2}}\Big(1-\frac{m_{H}^{2}}{m_{N}^{2}}\Big)^{2} .
\label{widths}
\eeq
When the heavy neutrino mass is much larger than the masses of the SM gauge bosons and the Higgs boson, i.e., $m_N \gg m_W, m_Z, m_H$, the branching ratios tend to the simple asymptotic proportion
\begin{equation}
\mathrm{BR}(N \to \ell W) : \mathrm{BR}(N \to \nu Z) : \mathrm{BR}(N \to \nu H) \simeq 2 : 1 : 1 .
\end{equation}

\section{Event Generation and Sensitivity Prospects\label{sec:analysis}}
We adopt a minimal benchmark scenario with a single phenomenologically
accessible heavy Majorana neutrino, denoted by $N\equiv N_1$.
We restrict the analysis to the muon-flavor-specific scenario,
\begin{equation}
|V_{\mu N}|^2 \neq 0, \qquad
|V_{eN}|^2 = |V_{\tau N}|^2 = 0.
\end{equation}
This choice avoids the stringent constraints from lepton-flavor-violating
(LFV) processes such as $\mu\to e\gamma$~\cite{MEG:2016leq}, which can arise
when electron and muon mixings are simultaneously nonzero, and allows us to
probe the dependence on $V_{\mu N}$ in a well-defined benchmark scenario.
The remaining two heavy neutrinos, $N_{2,3}$, are effectively decoupled by
setting their mixing parameters to zero and their masses to 10~TeV.

We consider the single production of $N$ at a future $\mu p$ collider with
\begin{equation}
\sqrt{s}=5.3~\text{TeV}, \qquad
E_p=7~\text{TeV}, \qquad
E_\mu=1~\text{TeV}.
\end{equation}

\begin{figure}[h]
\centering
\vspace{-2cm}
\includegraphics[width = 16cm ]{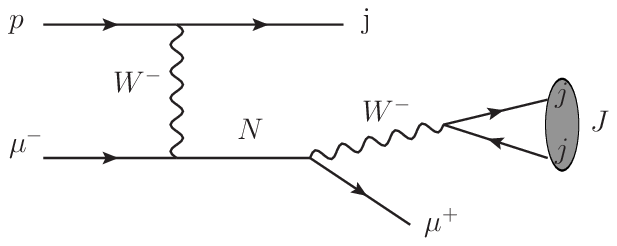}
\vspace{-18cm}
\caption{Feynman diagram for the LNV process
$\mu^-p\to jN$, followed by
$N\to\mu^+W^-\to\mu^+J$, at a $\mu p$ collider.
Here, $J$ denotes the fat jet originating from the hadronic decay of the
highly boosted $W$ boson. }
\label{fig1}
\end{figure}

For TeV-scale Majorana neutrinos, the $W$ boson produced in the decay
$N\to\mu^+W^-$ is highly boosted, such that its hadronic decay products
are typically reconstructed as a single fat jet $J$.
Together with the associated light jet $j$ from heavy-neutrino production,
the characteristic signal topology is therefore
\begin{equation}
\mu^-p\to jN, \qquad
N\to\mu^+W^-\to\mu^+J,
\end{equation}
with an observable final state $j\mu^+J$, as illustrated in
Fig.~\ref{fig1}. The charge-flipped final-state muon corresponds to
lepton-number violation by two units, $|\Delta L|=2$, and therefore provides
a direct probe of the Majorana nature of the heavy neutrino.

\begin{figure}[thb]
\begin{center}
\vspace{-0.5cm}
\centerline{\epsfxsize=10cm \epsffile{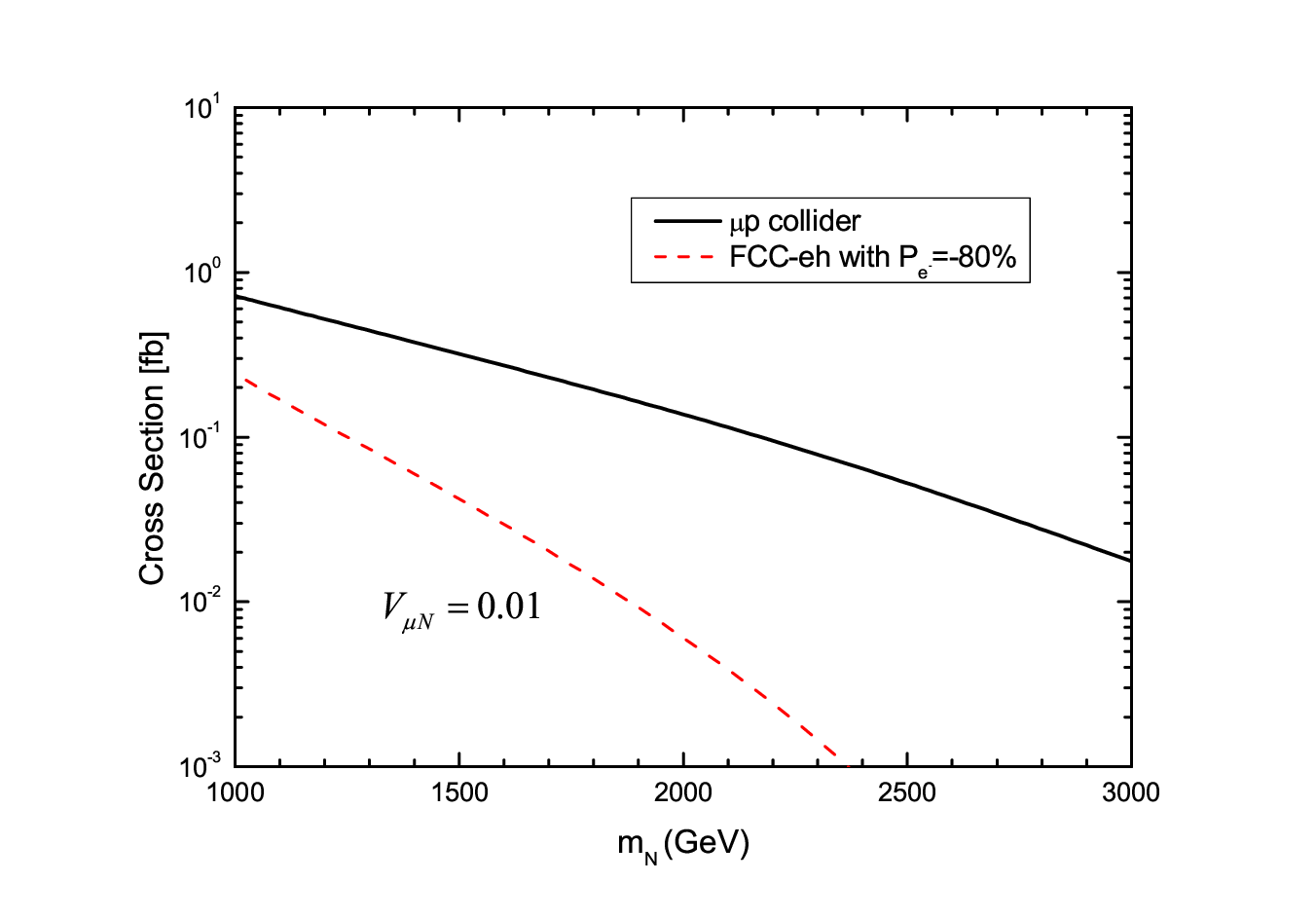}}
\caption{Cross sections for the LNV signal processes as a function of
$m_N$. The $\mu p$ result corresponds to
$\mu^-p\to jN\to j\mu^+J$ at $\sqrt{s}=5.3$~TeV in the
muon-flavor-specific scenario with $|V_{\mu N}|^2=10^{-4}$ and
$|V_{eN}|^2=|V_{\tau N}|^2=0$. For comparison, the FCC-eh result at
$\sqrt{s}=3.46$~TeV is taken from Ref.~\cite{Shen:2025ptk}, where
$|V_{eN}|^2=|V_{\mu N}|^2=10^{-4}$ is assumed.}
\label{fig2}
\end{center}
\end{figure}

Figure~\ref{fig2} presents the cross section for the LNV signal process
$\mu^-p\to jN\to j\mu^+J$ as a function of the heavy-neutrino mass at
the $\mu p$ collider, assuming the muon-flavor-specific benchmark
$|V_{\mu N}|^2=10^{-4}$ with
$|V_{eN}|^2=|V_{\tau N}|^2=0$.

For comparison, we also show the corresponding FCC-eh result for
$e^-p\to jN\to j\mu^+J$ at $\sqrt{s}=3.46$~TeV, obtained with a
60~GeV electron beam with $-80\%$ polarization and a 50~TeV proton
beam~\cite{Shen:2025ptk}. In Ref.~\cite{Shen:2025ptk}, the benchmark
choice $|V_{eN}|^2=|V_{\mu N}|^2=10^{-4}$ is adopted. Therefore, the
FCC-eh curve is shown here as a reference based on the flavor-mixing
scenario assumed in that study, rather than as a comparison under an
identical flavor benchmark.

For $m_N=2000$~GeV, the signal cross section at the $\mu p$ collider
is approximately $0.14$~fb, while the corresponding FCC-eh result is
$6.6\times10^{-3}$~fb. The cross section decreases with increasing
$m_N$ because of the reduced phase space and the suppression of the
relevant parton luminosities at larger momentum fractions.

The characteristic signal final state consists of a high-$p_T$
positively charged muon, a fat jet originating from the boosted hadronic
$W$ decay, and an associated light jet. Several Standard Model processes
can mimic this topology and therefore constitute relevant backgrounds.
Table~\ref{sm} lists the dominant background processes and their production
cross sections at the $\mu p$ collider. All cross sections are computed
at leading order using \textsc{MadGraph5\_aMC@NLO}~\cite{Alwall:2014hca}
with the NNPDF23L01 parton distribution functions~\cite{NNPDF:2014otw}.

\begin{table}[htb]
\centering
\caption{Dominant SM background processes and their cross sections at the $\mu p$ collider. Here $\ell$ denotes either an electron or a muon ($\ell = e, \mu$), and $q\bar{q}$ includes both light-flavor and $b\bar{b}$ final states.}
\label{sm}
\vspace{0.5em}
\begin{tabular}{lcc}
\toprule[1.5pt]
Process & Decay channel & $\sigma$ [fb]\\ \hline
$\mu^{-}p \to \mu^{+} \mu^{-} j \nu$ &
---&1054\\
$\mu^{-}p \to \nu_{\mu} W^{+} W^{-} j$ &
$W^{+}\to \ell^{+}\nu_{\ell}$, $W^{-}\to jj$ &
17.4 \\
$\mu^{-}p \to \mu^{-} W^{+} W^{-} j$ &
$W^{+}\to \ell^{+}\nu_{\ell}$, $W^{-}\to jj$ &
15.9 \\
$\mu^{-}p \to \mu^{-} W^{+} Z j$ &
$W^{+}\to \ell^{+}\nu_{\ell}$, $Z\to q\bar{q}$ &
4.1 \\
$\mu^{-}p \to \nu_{\mu} W^{+} Z j$ &
$W^{+}\to \ell^{+}\nu_{\ell}$, $Z\to q\bar{q}$ &
3.3\\
\hline
\end{tabular}
\end{table}

Signal events are generated using the Universal FeynRules Output (UFO)
implementation of the heavy-neutrino model~\cite{Alloul:2013bka}.
Parton-level events are generated with
\textsc{MadGraph5\_aMC@NLO} and subsequently passed through the simulation
chain. Parton showering and hadronization are simulated with
\textsc{Pythia~6}~\cite{Sjostrand:2006za}, while detector effects are modeled
using the fast detector simulation \textsc{Delphes~3.4.2}
~\cite{deFavereau:2013fsa} with a dedicated detector card for the
$\mu p$ collider. Fat jets are reconstructed with the Cambridge--Aachen
algorithm~\cite{Dokshitzer:1997in,Wobisch:1998wt}, as implemented in
\textsc{FastJet}~\cite{Cacciari:2011ma}, using a radius parameter $R=1.0$.
Event selection and analysis are finally performed within the
\textsc{MadAnalysis~5} framework~\cite{Conte:2012fm,Conte:2014zja}.

At the parton level, minimal acceptance requirements are applied to both signal and background samples:
\begin{equation}
p_{T}^{\ell/j} > 10~\text{GeV}, \qquad
|\eta_{\ell/j}| < 5, \qquad
\Delta R_{xy} > 0.4,
\end{equation}
where $p_{T}^{\ell/j}$ and $\eta_{\ell/j}$ represent the transverse momentum and pseudorapidity of leptons and jets, respectively, and $\Delta R_{xy} = \sqrt{\Delta\phi^{2}+\Delta\eta^{2}}$ defines the angular separation between any two objects $x$ and $y$.

\begin{figure*}[htb]
\begin{center}
\centerline{\hspace{2.0cm}\epsfxsize=9cm\epsffile{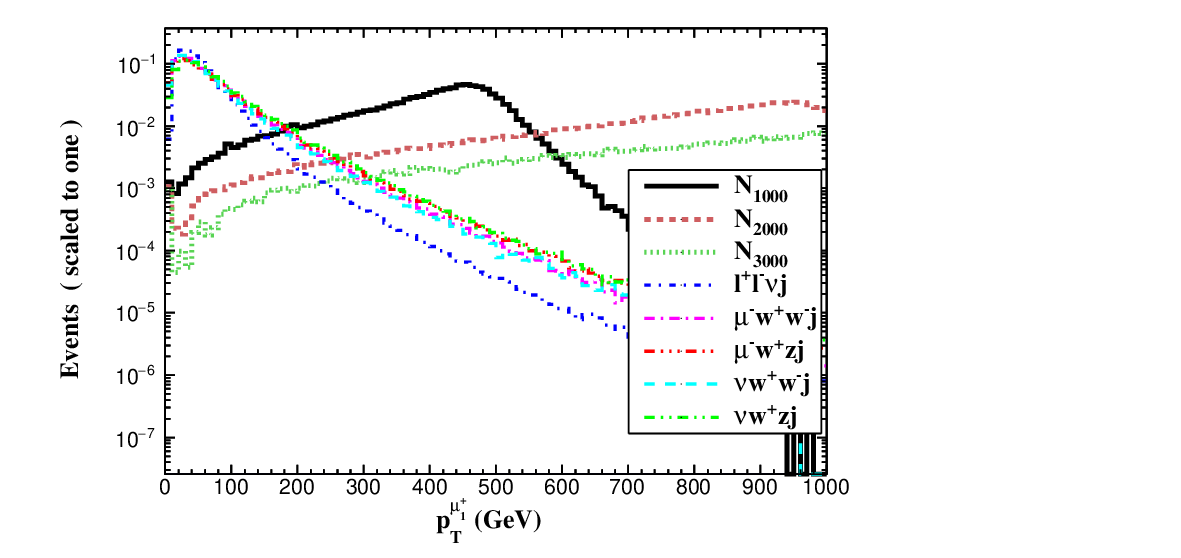}
\hspace{-2.0cm}\epsfxsize=9cm\epsffile{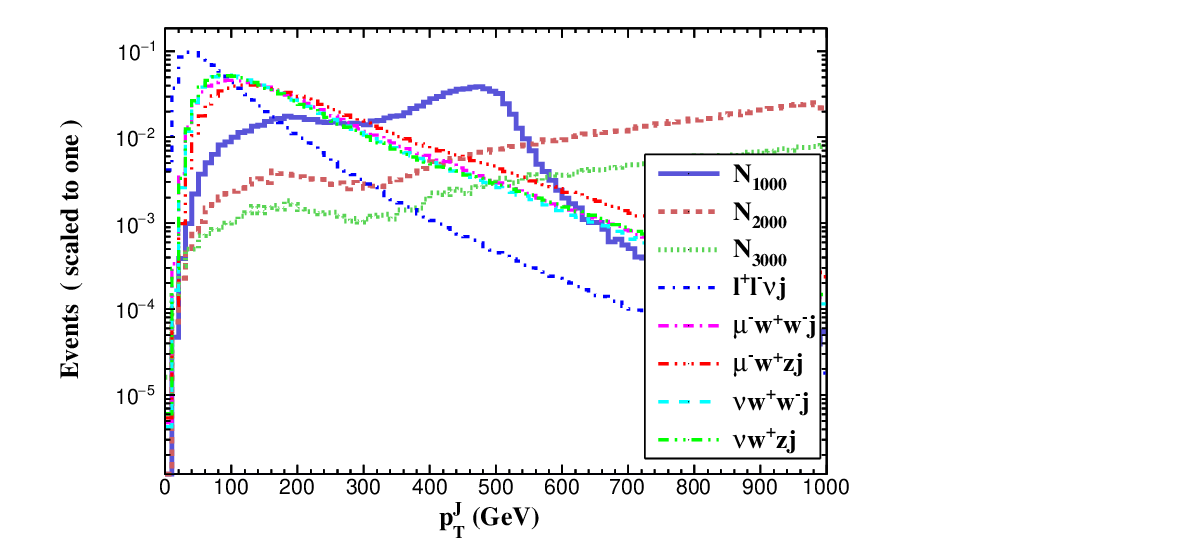}}
\centerline{\hspace{2.0cm}\epsfxsize=9cm\epsffile{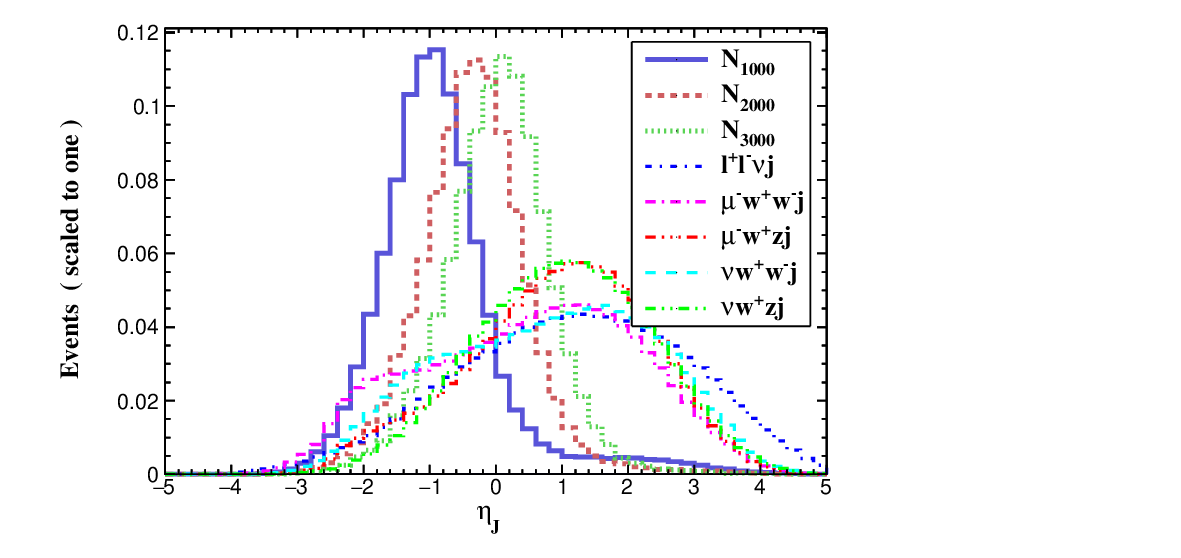}
\hspace{-2.0cm}\epsfxsize=9cm\epsffile{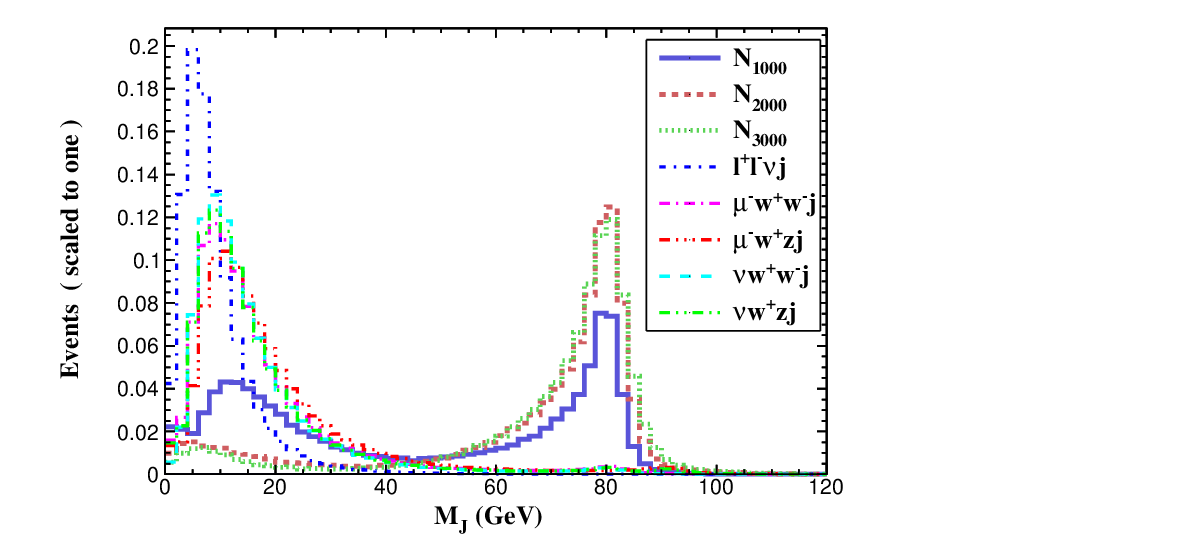}}
\centerline{\hspace{2.0cm}\epsfxsize=9cm\epsffile{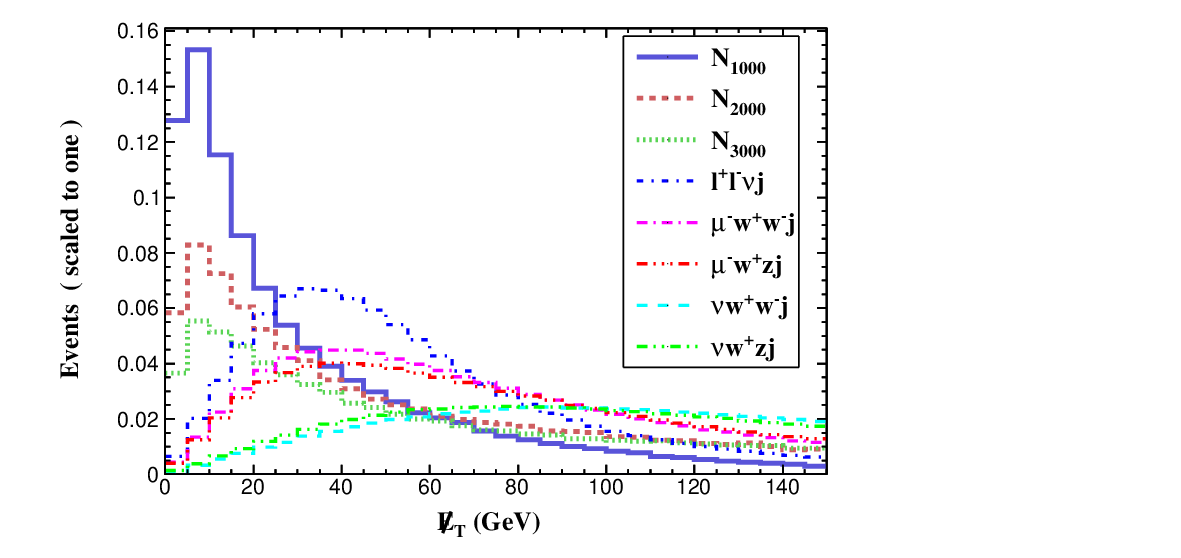}
\hspace{-2.0cm}\epsfxsize=9cm\epsffile{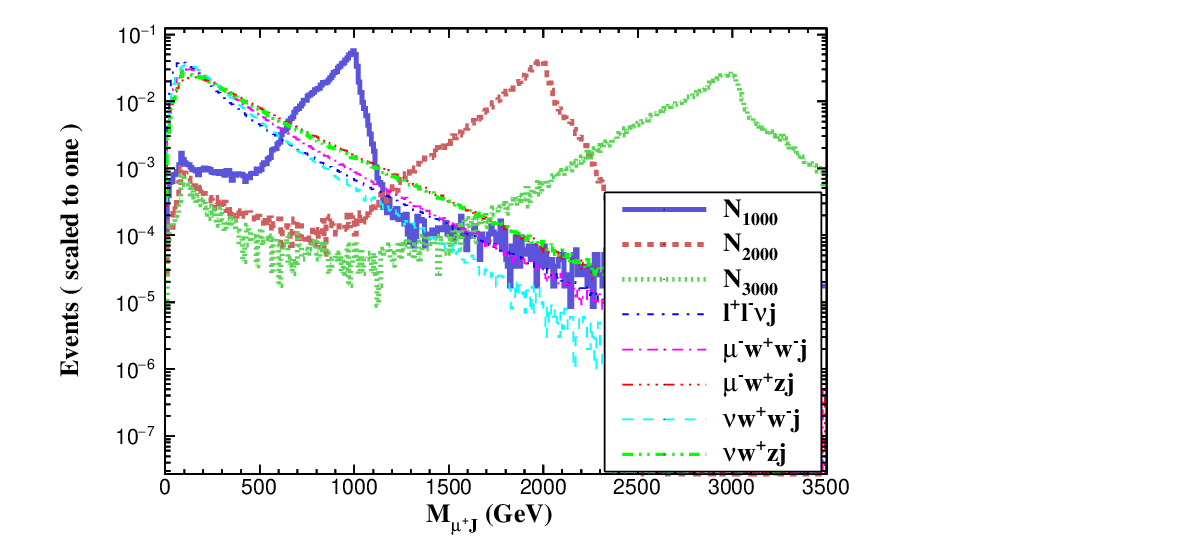}}
\caption{Normalized distributions for the signals (with $m_{N}=1000, 2000$ and 3000~GeV) and SM backgrounds at the $\mu p$ collider, including the transverse momentum of the muon $p_T^{\mu^+}$, the transverse momentum of the fat-jet $p_T^J$, the pseudorapidity of the fat-jet $\eta_J$, the fat-jet mass $M_J$, the missing transverse energy $\slashed{E}_T$, and the invariant mass of the muon-fat-jet system $M_{\mu^+ J}$.}
\label{fig3}
\end{center}
\end{figure*}

Figure~\ref{fig3} shows the normalized kinematic distributions for the
signal benchmarks and the SM backgrounds at the $\mu p$ collider.
Because of the large heavy-neutrino mass, the signal exhibits substantially
harder $p_T^{\mu^+}$ and $p_T^J$ spectra than the SM backgrounds.
The background distributions also extend more strongly toward larger
$\slashed{E}_T$, whereas the signal events are concentrated in the
low-$\slashed{E}_T$ region. In addition, the signal distribution of
$M_{\mu^+J}$ is concentrated around the corresponding heavy-neutrino mass,
making this variable particularly effective for signal-background
discrimination.

To optimize signal extraction from background processes, we implement a series of sequential kinematic cuts based on the distinctive features observed in the differential distributions:
\begin{itemize}
    \item Cut 1: Exactly one positively charged muon,
    $N(\mu^+)=1$, is required, with
    $p_T^{\mu^+}>150~\text{GeV}$ and
    $|\eta_{\mu^+}|<2.5$.

    \item Cut 2: At least one fat-jet candidate $J$ and one additional
    light jet $j$ are required. Their transverse momenta must satisfy
    $p_T^J>200~\text{GeV}$ and $p_T^j>20~\text{GeV}$, respectively,
    with $|\eta_J|<2.5$.

    \item Cut 3: The fat-jet mass is required to be consistent with a
    hadronically decaying $W$ boson,
    $|M_J-m_W|<15~\text{GeV}$.

    \item Cut 4: The missing transverse energy is required to satisfy
    $\slashed{E}_T<40~\text{GeV}$.

    \item Cut 5: The invariant mass of the muon--fat-jet system is required
    to satisfy $M_{\mu^+J}>0.9m_N$.
\end{itemize}

\begin{table}[htb]
\centering
\caption{Cut flow of the cross sections (in fb) for signal benchmarks with $m_N = 1000, 2000, 3000~\text{GeV}$ and for the dominant SM backgrounds at the $\mu p$ collider. The mixing parameter is set to $|V_{\mu N}|^{2}=10^{-4}$. Since Cut~5 depends explicitly on $m_N$, the three Cut~5 rows correspond
to the mass-dependent selections for $m_N=1$, 2, and 3~TeV, respectively.\label{cutflow2}}
\vspace{0.5em}
\begin{tabular}{@{}lccccccccc@{}}
\toprule[1.5pt]
\multirow{2}{*}{Cuts} & \multicolumn{3}{c}{Signal (fb)} & & \multicolumn{5}{c}{Background (fb)} \\
\cline{2-4} \cline{6-10}
& 1000~GeV & 2000~GeV & 3000~GeV & & $\mu^{-}\mu^{+}j\nu$ & $\mu^{-}W^{+}W^{-}j$ & $\mu^{-}W^{+}Zj$ & $\nu W^{+}W^{-}j$ & $\nu W^{+}Zj$ \\ \cline{1-10}
\midrule[1pt]
Basic & 0.68 & 0.13 & 0.016 & & 864 & 15.3 & 3.89 & 11.48 & 3.07 \\
Cut 1 & 0.63 & 0.12 & 0.015 & & 5.69 & 0.044 & 0.011 & 0.57 & 0.018 \\
Cut 2 & 0.51 & 0.11 & 0.014 & & 1.58 & 0.032 & 0.009 & 0.27 & 0.009 \\
Cut 3 & 0.43 & 0.098 & 0.013 & & 0.038 & 0.0031 & 0.0009 & 0.032 & 0.001 \\
Cut 4 & 0.17 & 0.039 & 0.0051 & & 0.0028 & $3.1\times10^{-4}$ & $1.1\times10^{-4}$ & $6.1\times10^{-4}$ & $2.4\times10^{-5}$ \\
\multirow{3}{*}{Cut 5} & 0.162 & -- & -- & & 0.0014 & $1.1\times10^{-4}$ & $3.7\times10^{-5}$ & $2.7\times10^{-4}$ & $1.7\times10^{-5}$ \\
& -- & 0.035 & -- & & $2.85\times10^{-4}$ & $3.2\times10^{-6}$ & $1.6\times10^{-6}$ & $3.0\times10^{-5}$ & $1.3\times10^{-6}$ \\
& -- & -- & 0.0046 & & $1.4\times10^{-4}$ & $1.2\times10^{-6}$ & $8.3\times10^{-7}$ & $1.2\times10^{-5}$ & $5.4\times10^{-7}$ \\
\bottomrule[1.5pt]
\end{tabular}
\end{table}

The cross sections for the three representative signal benchmarks,
$m_N=1000$, 2000, and 3000~GeV, together with the corresponding SM
backgrounds, are summarized in Table~\ref{cutflow2}.
The sequential selection requirements strongly suppress all of the
background processes while retaining an appreciable fraction of the signal.
After the complete selection, the dominant residual background arises from
the $\mu^-p\to\mu^+\mu^-j\nu$ process. The total cross section of SM backgrounds is approximately $1.86\times10^{-3}$~fb for $m_N=1000$~GeV, reducing to $3.2\times10^{-4}$~fb for $m_N=2000$~GeV and $1.5\times10^{-4}$~fb for $m_N=3000$~GeV.

Among the SM background processes, particular attention is paid to the contribution from quasi-real photon exchange, which can produce a collinear muon and mimic the signal topology. Such processes are simulated by including photon-initiated channels, $p\gamma\to \mu^{+}jjj\nu$, via the photon PDFs of the proton. After applying the full set of selection cuts, we find that the residual cross sections from these photon-induced backgrounds are $1.2\times10^{-4}$~fb, $3.4\times10^{-6}$~fb, and $1.2\times10^{-6}$~fb for $m_N = 1$, $2$, and $3$~TeV, respectively, corresponding to approximately $6.7\%$, $1\%$, and $0.7\%$ of the total SM background. These contributions remain at the percent level and thus have a negligible impact on the final sensitivity.

To quantify the projected sensitivity, we use the exclusion significance
\begin{equation}
\mathcal{Z}_{\rm exc}
=
\sqrt{2\left[
s-b\ln\left(1+\frac{s}{b}\right)
\right]},
\end{equation}
where $s$ and $b$ denote the expected numbers of signal and background
events, respectively~\cite{Cowan:2010js}.
Systematic uncertainties are not included in the present projection.

\begin{figure}[ht]
\begin{center}
\vspace{1.5cm}
\centerline{\epsfxsize=12cm \epsffile{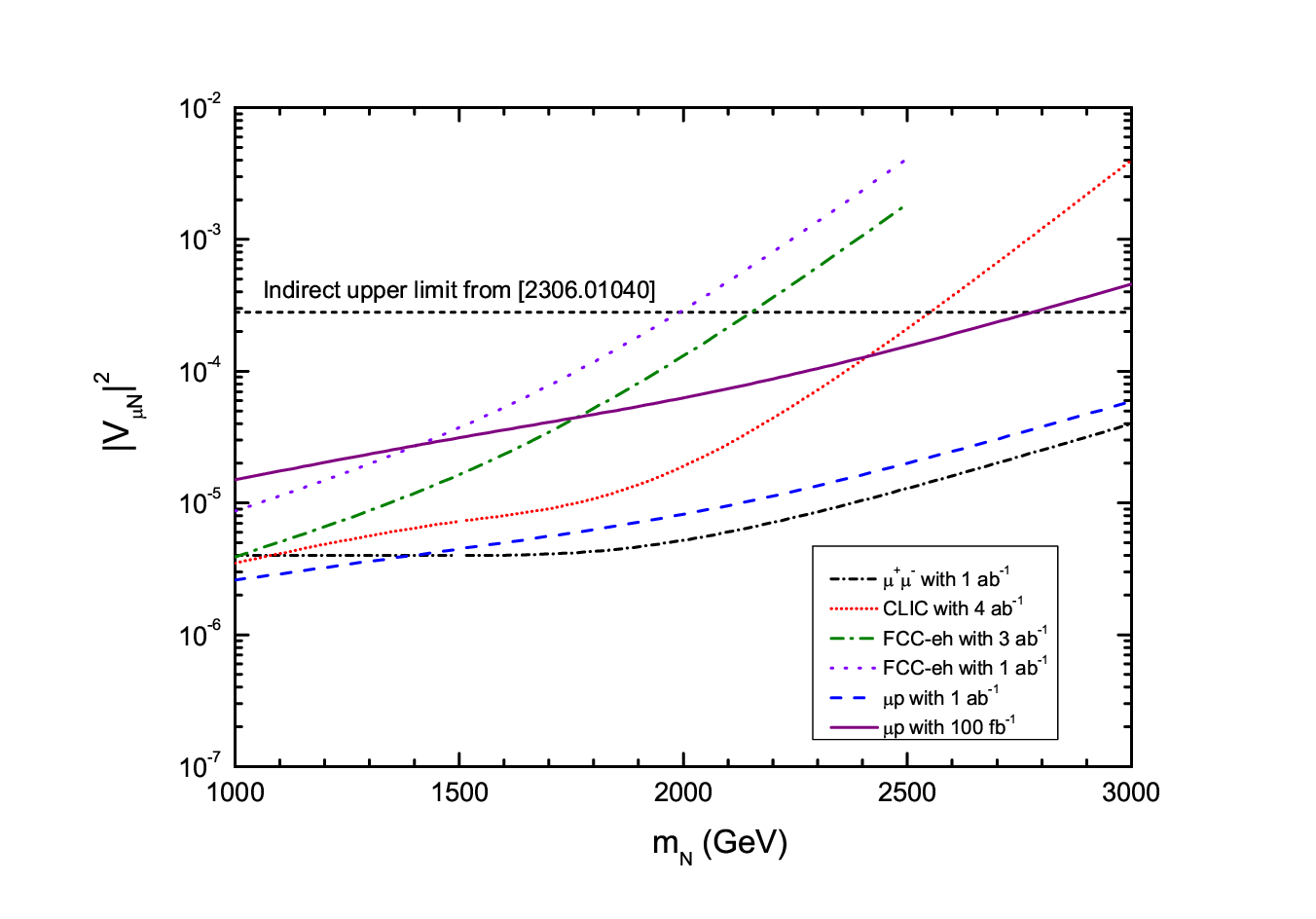}}
\caption{Projected $2\sigma$ exclusion limits in the
$|V_{\mu N}|^2$--$m_N$ plane for the $j\mu^+J$ signal at the
$\mu p$ collider. Also shown are projected sensitivities from CLIC,
a 3~TeV muon collider, and the FCC-eh, together with the updated
indirect bound from Ref.~\cite{Blennow:2023mqx}. The comparison curves
are shown using the flavor-mixing assumptions adopted in the respective
references; in particular, the FCC-eh result of
Ref.~\cite{Shen:2025ptk} assumes
$|V_{eN}|^2=|V_{\mu N}|^2$.}
\label{fig4}
\end{center}
\end{figure}

In Fig.~\ref{fig4}, we present the projected $2\sigma$ exclusion limits
in the $|V_{\mu N}|^2$--$m_N$ plane for the muon-flavor-specific scenario,
$|V_{\mu N}|^2\neq0$ and
$|V_{eN}|^2=|V_{\tau N}|^2=0$.
Results are shown for integrated luminosities of
$100~\mathrm{fb}^{-1}$ and $1~\mathrm{ab}^{-1}$.
For comparison, we also show the updated global indirect constraint on
$|V_{\mu N}|^2$ from Ref.~\cite{Blennow:2023mqx}.

Even at $100~\mathrm{fb}^{-1}$, the projected $\mu p$ sensitivity improves
upon the indirect constraint over most of the mass range considered.
At $1~\mathrm{ab}^{-1}$, the improvement reaches approximately one order
of magnitude in the lower-mass region.
As $m_N$ increases from 1 to 3~TeV, the projected $2\sigma$ upper limit on
$|V_{\mu N}|^2$ varies from
$1.54\times10^{-5}$ to $4.6\times10^{-4}$ for
$\mathcal{L}=100~\mathrm{fb}^{-1}$,
and from $2.6\times10^{-6}$ to $5.9\times10^{-5}$ for
$\mathcal{L}=1~\mathrm{ab}^{-1}$.

In Ref.~\cite{Mekala:2022cmm}, the authors investigated heavy Majorana
neutrinos in the mass range $200$--$3200$~GeV via the $qq\ell$ final state
at the 3~TeV CLIC, assuming an integrated luminosity of
$4~\mathrm{ab}^{-1}$ and an electron beam polarization of $-80\%$.
Reference~\cite{Li:2023tbx} explored the sensitivity of a muon collider
operating at $\sqrt{s}=3$~TeV with an integrated luminosity of
$1~\mathrm{ab}^{-1}$ to heavy neutral leptons.
In Ref.~\cite{Shen:2025ptk}, the authors studied the LNV process
$e^-p\to jN\to j\mu^+J$ at the FCC-eh with 60~GeV electrons and
50~TeV protons, considering heavy-neutrino masses in the range
$1000$--$2500$~GeV and integrated luminosities of
$1$ and $3~\mathrm{ab}^{-1}$.
In that analysis, the correlated two-flavor benchmark
$|V_{eN}|^2=|V_{\mu N}|^2$ was assumed. Therefore, the FCC-eh sensitivity
shown for comparison should be interpreted within this correlated
two-flavor scenario, rather than as a model-independent constraint on
$|V_{\mu N}|^2$ alone.
Taken together, these comparisons show that a future $\mu p$ collider
offers complementary sensitivity to TeV-scale heavy Majorana neutrinos
relative to other proposed high-energy facilities.

\section{Conclusion\label{sec:summary}}

In this work, we have investigated the sensitivity to TeV-scale heavy
Majorana neutrinos at a future muon-proton collider with
$\sqrt{s}=5.3$~TeV, based on 1~TeV muon and 7~TeV proton beams.
We consider a muon-flavor-specific benchmark with
$|V_{\mu N}|^2\neq0$ and $|V_{eN}|^2=|V_{\tau N}|^2=0$, and focus on the
LNV process $\mu^-p\to jN$, followed by
$N\to\mu^+W^-\to\mu^+J$, which leads to the characteristic
$j\mu^+J$ final state. For TeV-scale heavy neutrinos, the $W$ boson
produced in the heavy-neutrino decay is highly boosted, such that its
hadronic decay products can be efficiently reconstructed as a single
fat jet. This characteristic topology enables an effective suppression
of the SM backgrounds.

After applying the optimized event-selection criteria, we derive the
projected $2\sigma$ exclusion limits on the muon-flavor mixing parameter
$|V_{\mu N}|^2$. For heavy-neutrino masses in the range
$1~\mathrm{TeV}\leq m_N\leq3~\mathrm{TeV}$, the projected limits vary from
$1.54\times10^{-5}$ to $4.6\times10^{-4}$ for an integrated luminosity of
$100~\mathrm{fb}^{-1}$, and from
$2.6\times10^{-6}$ to $5.9\times10^{-5}$ for
$1~\mathrm{ab}^{-1}$.
These results indicate that a future $\mu p$ collider can provide
competitive and complementary sensitivity to TeV-scale heavy Majorana
neutrinos, offering an additional probe alongside other proposed
high-energy collider facilities.

\begin{acknowledgments}
The work is supported by the Natural Science Foundation of Henan Province~(Grant No.~252300421988).
\end{acknowledgments}

\end{document}